\documentclass[%
reprint,
superscriptaddress,
nofootinbib,
 amsmath,amssymb,
 aps,
prb,
]{revtex4-2}

\usepackage[normalem]{ulem}
\usepackage{graphicx}
\usepackage{dcolumn}
\usepackage{bm}
\usepackage{hyperref} 
\usepackage{xcolor}

\usepackage[makeroom]{cancel}

\begin{document}

\preprint{APS/123-QED}

\title{Hydrodynamics of dipole-conserving fluids}

\author{Aleksander G\l{}\'{o}dkowski}
\affiliation{Institute for Theoretical Physics, Wroc\l{}aw  University  of  Science  and  Technology,  50-370  Wroc\l{}aw,  Poland}
\affiliation{Max Planck Institute for the Physics of Complex Systems, 01187 Dresden, Germany}
\author{Francisco Pe\~na-Ben\'itez}
\affiliation{Institute for Theoretical Physics, Wroc\l{}aw  University  of  Science  and  Technology,  50-370  Wroc\l{}aw,  Poland}
\author{Piotr Sur\'{o}wka}%
\affiliation{Institute for Theoretical Physics, Wroc\l{}aw  University  of  Science  and  Technology,  50-370  Wroc\l{}aw,  Poland}
\affiliation{Max Planck Institute for the Physics of Complex Systems, 01187 Dresden, Germany}

\date{\today}

\begin{abstract}
Dipole-conserving fluids serve as examples of kinematically constrained systems that can be understood on the basis of symmetry. They are known to display various exotic features including glassylike dynamics, subdiffusive transport, and immobile excitations' dubbed fractons. Unfortunately, such systems have so far escaped a complete macroscopic formulation as viscous fluids. In this work, we construct a consistent hydrodynamic description for fluids invariant under translation, rotation, and dipole shift symmetry. We use symmetry principles to formulate a thermodynamic theory for dipole-conserving systems at equilibrium and apply irreversible thermodynamics in order to elucidate dissipative effects. Remarkably, we find that the inclusion of the energy conservation not only renders the longitudinal modes diffusive rather than subdiffusive but also diffusion is present even at the lowest order in the derivative expansion. This work paves the way towards an effective description of many-body systems with constrained dynamics such as ensembles of topological defects, fracton phases of matter, and certain models of glasses.
\end{abstract}

\maketitle


\section{Introduction}
Over the years, hydrodynamics has evolved into a universal framework describing the long-wavelength dynamics of many-body systems. It provides a systematic scheme for the evolution of conserved charges, leading to an effective description that is for the most part irrespective of the microscopic details. The structure of a hydrodynamic theory is determined by the symmetries of the system at hand, once the appropriate low-energy degrees of freedom have been identified and a finite set of phenomenological transport coefficients introduced. This approach gives access to the low-energy (long-wavelength) physics of complex many-body systems that are often impossible to be understood from first principles. Over time, the hydrodynamic paradigm has proven to be a robust tool in describing both classical and quantum liquids \cite{PhysRevB.51.13389,Shuryak_2009,Cao_2011,Crossno_2016}. Recent effort to apply the formalism of hydrodynamics concentrates, among others, on systems with kinematic constraints. A variant of this problem is particularly important in the field of amorphous solids or glasses \cite{berthier_dynamical_2011}.

Classical glasses refer to any noncrystalline solid that exhibits a glass transition when heated towards the liquid state. Models of classical glasses are largely stochastic lattice models with imposed kinetic constraints on the allowed transitions between different configurations of the system, while preserving equilibrium conditions. The glassy behavior of such systems is visible in the equilibration timescales that grow exponentially with system size. Unfortunately, the constrained models proposed to elucidate classical glasses are only subject to numerical studies and have not been systematically analyzed in terms of the low-energy hydrodynamic theory.  

Supercooled liquids are not the only physical systems with glassy characteristics where kinematically constrained models appear. Such models emerge in many-body ensembles of elastic defects \cite{duality,Gromov2020,Radzihovsky_2020,surowka2021,Caddeo:2022ibe}, vortices in superfluids \cite{doshi_vortices_2021,Nguyen_2020}, or dimer-plaquette models \cite{You_Fractonic_2022,PhysRevB.103.094303}. Several quantum systems, such as lattice models \cite{Chamon_Quantum_2005,Haah_Local_2011,Bravyi_2011,PhysRevB.97.165106,Myerson_Construction_2022},  spin liquids \cite{xu_gapless_2006,Pretko_2017,Pretkoo_2017,Yan_2020,Benton_2021}, quantum bosonic matter \cite{Sandvik_striped_2002,Rousseau_Phase_2004,Giergiel_BH_2021,Lake_Dipolar_2022,Lake:2022hel,Zechmann:2022esq}, and tilted optical lattices in external magnetic fields \cite{Guardado_Sanchez_2020,scherg_observing_2021} have also been shown to exhibit mobility restrictions. In addition, kinematic constraints play an important role in the novel fractonic phases of matter that are yet to be realized \cite{Vijay_2015}. A detailed  understanding of the low-energy dynamics is therefore crucial for future experimental proposals and transport measurements.

Systems with mobility restrictions inspired a vast amount of work in theoretical physics, e.g. see \cite{Grosvenor:2021hkn,Gromov:2022cxa} and references therein. Recent developments revealed that theories conserving multipole moments of conserved charges go beyond conventional field theories \cite{gromov_towards_2019,Seiberg:2020bhn,Gorantla:2021svj,Gorantla__low_2021}, requiring a different approach for integrating high-energy modes \cite{Lake_Renormalization_2022,Grosvenor:2022ohr}, and leading to new geometric structures that appear when coupled to background geometries \cite{francisco,Bidussi_2022,Jain_2022}. Consequently, transport properties of kinematically constrained many-body systems are fundamentally different from theories without mobility restrictions. This is manifested in the thermalization properties of systems with emergent dipole-conservation that exhibit a characteristic subdiffusive behavior \cite{Feldmeier_anomalous_2020,Morningstar_kinetically_2020,Iaconis_Multipole_2021,Moudgalya_spectral_2021}. Such systems have been studied both experimentally \cite{Guardado_Sanchez_2020,scherg_observing_2021} and numerically \cite{Morningstar_kinetically_2020,Feldmeier_anomalous_2020}. These developments have lead to theoretical progress aiming at a consistent hydrodynamic theory of fluids with dipole moment conservation, which is sometimes referred to as \textit{fracton hydrodynamics}. 

First, the hydrodynamic theory for simple systems conserving charge and dipole, but not momentum, has been proposed \cite{Gromov_hydro_2020}. Subsequently, ideal hydrodynamics for several classes of kinematically constrained fluids with momentum conservation was formulated \cite{Grosvenor_hydrodynamics_2021}. Finally, dissipative effects for fracton fluids without energy conservation were studied in Ref. \cite{GloriosoLucas22} and other works followed thereafter \cite{Osborne_infinite_2022,PhysRevE.105.044103,PhysRevLett.129.150603,PhysRevB.105.205127}.

In addition to filling in some pedagogical gaps, we strive to generalize previous work by providing a systematic treatment of dissipative dipole-conserving fluids with both, momentum and energy conservation, using the method of irreversible thermodynamics. Within our theory, momentum per particle behaves as a St\"ueckelberg field, therefore, in analogy with the superfluid, we propose that constant gradients of such a variable are allowed at thermodynamic equilibrium. In fact, the conjugate variable to the gradient of momentum is understood as the flux of dipoles. Our main results are a set of dynamical equations for zeroth (Eqs. \eqref{eq:linearziedEoms}) and first order (Eqs. \eqref{eq:eoms1}) linear hydrodynamics, respectively. Contrary to the case of ideal ordinary fluids, the zeroth order equations of motion contain a dissipative transport coefficient interpreted as a thermal conductivity. On the other hand, the first order hydrodynamic equations contain 1+12 transport coefficients. Strikingly, the thermal conductivity $\alpha$ modifies the structure of the hydrodynamic modes predicted in \cite{Grosvenor:2021hkn,GloriosoLucas22}, upgrading the longitudinal modes to an ordinary diffusive form $\omega_{||}\sim -i\alpha k^2$, while allowing for subdiffusion only in the shear sector $\omega_{shear} \sim-i\eta k^4$, where $\eta$ is the analogue to the shear viscosity of the system. In addition, for some regions in the parameter space, the soundlike modes scale as $\pm k^2-ik^2$, whereas in the complementary regions they scale as $-ik^2$. Therefore, their propagation will be either strongly attenuated or fully diffusive.

The paper is organized as follows. In Sec. \ref{sec:symmetries} we introduce the symmetries that characterize the kinematically constrained fluids studied in this work. 
In Sec. \ref{sec:thermodynamics} we propose a thermodynamic description compatible with the symmetries. In Sec. \ref{sec:hydrodynamics} the hydrodynamic expansion is systematically derived on the basis of the entropy current formalism, the constitutive relations are determined and the hydrodynamic modes are studied. Finally, in Sec. \ref{sec:discussion} we conclude with a brief discussion and outlook.

\section{Symmetries}\label{sec:symmetries}
Let us start by introducing the set of symmetries underlying our hydrodynamic construction of kinematically constraint fluids.

To this end, we consider a many-body system enjoying a simultaneous conservation of energy $\mathcal{H}$, momentum $\mathcal P_i$, U(1) charge $\mathcal Q$, dipole moment $\mathcal D_i$ and angular momentum $\mathcal J_{ij}$. In the long-wavelength regime, the conserved charges can be expressed in terms of the local densities 
\begin{equation}\label{eq:conserved}
\begin{split}
     \mathcal H   &= \int d^d x \,\epsilon \,, \quad \mathcal P_i  = \int d^d x \,p_i \,,  \\
 \mathcal Q   &= \int d^d x \,n\,, \quad \mathcal D_i  = \int d^d x \hspace{2pt}x_i n \,, \\
 \mathcal J_{ij}  &=  \int d^d x \,\big( x_i p_j  - x_j p_i  \big)\,.
\end{split}
\end{equation}
Furthermore, we will impose parity and time-reversal invariance. The long-wavelength dynamics near thermodynamic equilibrium will be governed by the gapless degrees of freedom. Of course, it is only the locally conserved densities that remain relevant, as non-conserved quantities are expected to be fast-relaxing. Therefore, the hydrodynamic equations will be the local conservation laws
\begin{equation}
\label{eq:continuity}
\begin{split}
 \partial_t  n &+ \partial_i \partial_j J^{ij} = 0\,,\\
  \partial_t  p_i &+ \partial_j T^{ji} = 0\,,\\
\partial_t  \epsilon  &+ \partial_i J^i_\epsilon = 0\,.
\end{split}
\end{equation}
Note that neither dipole nor angular conservation lead to additional hydrodynamic equations, since their conservation follows from Eqs. \eqref{eq:continuity}, provided that the stress tensor $T^{ij}$ is symmetric.

The set of transformations generating the conserved charges form a Lie group with algebra\footnote{Throughout the paper we will use squared brackets to refer to antisymmetrization of indices  $A_{[ij]} = 1/2(A_{ij}-A_{ji})$, and symmetrization with parenthesis $A_{(ij)} = 1/2(A_{ij}+A_{ji})$.} as follows:
\begin{equation}
\begin{split}
     \{ \mathcal J_{ij},  \mathcal J_{kl}\} &= 2\delta_{i[k}  \mathcal J_{l]j} + 2\delta_{j[l}  \mathcal J_{k]i}\,, \\
     \{ \mathcal J_{ij}, \mathcal P_k\} &=  2\delta_{k[i} \mathcal P_{j]}\,,\\
      \{ \mathcal J_{ij}, \mathcal D_k\} &=  2\delta_{k[i} \mathcal D_{j]}\,, \\
\{\mathcal D_{i}, \mathcal P_j\} &= \delta_{ij} \mathcal Q\,.
\end{split}
\label{eq:commutations}
\end{equation}
Notice that the last bracket in Eq. \eqref{eq:commutations} implies, that momentum is not invariant under the transformation generated by $\mathcal D_i$. In fact, it transforms as
\begin{equation}
    \delta_{\bm{\beta}}\mathcal  P_i = -\beta_j \mathcal Q\,.
\end{equation}
Consequently, the momentum density and the stress tensor must transform under dipole shift in the following manner:
\begin{equation}
\label{eq:transformation2}\delta_{\bm{\beta}}p_i = - n \beta_i,\,\,
\delta_{\bm{\beta}}T^{ij} = \partial_{k} J^{ij} \beta_{k} - \partial_{k} J^{kj} \beta_{i} -  \partial_{k} J^{ki} \beta_{j} \,.
\end{equation}
This transformation property was also obtained in Ref. \cite{Jain_2022} by placing the system in curved space and coupling it to Aristotelian background sources as well as appropriate gauge fields. As we see in the following sections, these unusual transformation properties lead to a number of exotic features in the hydrodynamics (and thermodynamics) of dipole-conserving fluids.

Finally, we note that algebra in Eqs. \eqref{eq:commutations} is incompatible with boost symmetry (Galilean or Lorentz). In fact, non-boost invariant fluids without dipole symmetry have been recently subjected to intensive theoretical analysis \cite{perfect,10.21468/SciPostPhys.5.2.014,hydrowithoutboost,effectivehydrowithoutboost,de_Boer_2020}.

\section{Thermodynamics and dipole conservation}\label{sec:thermodynamics}
Thermodynamics of dipole-conserving systems cannot be captured by the standard textbook treatment. It requires a modified approach that systematically incorporates kinematic constraints arising from the dipole conservation. In this section, we construct a consistent thermodynamic theory with dipole conservation built into it.
\subsection{Dipole-invariant equation of state}
The internal energy density of a generic system in equilibrium is a function of the entropy and conserved charge densities, for example $\epsilon \equiv \epsilon(n,s,p_i)$. However, owing to the noncommutative structure of the algebra \eqref{eq:commutations}, dipole-conserving systems are not generic. In fact, the combination $p_i/n$ has a shift symmetry under dipole transformations in analogy to a Nambu-Goldstone mode; therefore, it could enter via the invariant combination $V_{ij}=\partial_i(n^{-1} p_j)$. It is then necessary to introduce a conjugate variable $F_{ij}$ that, as we will see, can be interpreted as a flux of dipoles.
 Thus, we infer that different (constant) values of $V_{ij}$ label distinct thermodynamic states. For such systems, we postulate that the first law of thermodynamics reads\footnote{Since we are allowing for gradients of conserved quantities in the equilibrium state, we could think of this formulation as describing a hydrostatic regime.}
\begin{equation}
    d\epsilon = Tds + \mu dn + F_{ij}dV_{ij}\,.
\end{equation}
Besides rotational invariance, we have also assumed that microscopically the system preserves parity. After noticing that under parity transformations  $V_{[ij]}$ transform as a pseudo-scalar and as a pseudo vector in two and three dimensions respectively, we conclude the energy density must depend on the symmetric part of $V_{ij}$ only. 
The pressure of the system is then defined via the standard thermodynamic relation
\begin{equation}\label{eq:pressureDef}
    P = Ts + \mu n -\epsilon\,, \quad dP = n d\mu + s dT - F_{ij} d V_{(ij)}\,.
\end{equation}
From now on we will refer to the symmetrized tensor as $V_{ij}$. 
In our construction, we shall take $n,\epsilon,p_i$ as the hydrodynamic variables. Therefore, we find convenient to use entropy density as the thermodynamic potential. Within this picture,
\begin{equation}\label{eq:diffentropy}
    d s = \frac{1}{T} d \epsilon  - \frac{\mu}{T} dn -  \frac{F_{ij}}{T} d V_{ij}
\end{equation}
where we have defined thermodynamic quantities 
\begin{equation}
    \label{eq:thermodynamicsDefinitions}
\frac{1}{T} =  \frac{\partial s}{\partial \epsilon} \,, \qquad
\frac{\mu}{T} = -  \frac{\partial s}{\partial n}  \,,    \qquad
\frac{F_{ij}}{T} = - \frac{\partial s}{\partial V_{ij}} \,.
\end{equation}
The relation $s \equiv s(\epsilon, n, V_{ij})$ is then interpreted as the equation of state and the thermodynamic quantities are understood as functions of the dipole-invariant variables $(\epsilon, n, V_{ij})$.
\subsection{Thermodynamic relations}
In the next section, we will study linearized hydrodynamics around the global equilibrium state $(n=n_0,\epsilon=\epsilon_0,V_{ij}=0)$. Therefore, it will be useful to introduce a set of thermodynamic identities that will allow us to relate the variations of $(\epsilon,n,V_{ij})$ with their corresponding conjugate variables. 

To do so, we first expand the entropy density function around the equilibrium state up to the second order in fluctuations
\begin{align}
 \nonumber   s &=  s_0 - \frac{\mu_0}{T_0} \delta n +\frac{1}{T_0} \delta \epsilon+ \frac{1}{2}s_{nn} \delta n ^2 + \frac{1}{2}s_{\epsilon \epsilon} \delta \epsilon ^2 + s_{n\epsilon} \delta \epsilon \delta n \\
    & - \frac{1}{2T_0} f_{||} \delta V_{kk} ^2  - \frac{1}{2T_0} f_{\perp} \delta V_{\langle ij \rangle} ^2\,,
    \end{align}
where $s_0$ is the entropy evaluated at the equilibrium state, the traceless symmetrization defined as $A_{\langle i j \rangle}= \frac{1}{2}(A_{ij}+A_{ji}-\frac{2}{d}A_{kk}\delta_{ij})$ and $A_{kk} = \delta_{ik} A_{ik}$ denoting the trace. Thermodynamic stability imposes the constraints
\begin{equation}\label{eq:stability}
    f_{||},f_\perp\geq 0\,,\quad s_{\epsilon\epsilon},s_{nn} < 0 \,,\quad s_{n\epsilon}^2- s_{n n}s_{\epsilon\epsilon} \leq 0\,.
\end{equation}
Therefore, the variations of the thermodynamic quantities Eq. \eqref{eq:thermodynamicsDefinitions} can be expressed as
\begin{equation}\label{eq:therm1}
\begin{split}
     \delta  \frac{1}{T} &= s_{\epsilon \epsilon} \delta \epsilon +  s_{\epsilon n} \delta n \,, \\
     \delta  \frac{\mu}{T} &= - s_{n \epsilon} \delta \epsilon -  s_{n n} \delta n \,, \\
     \delta  F_{ij} &=  f_{||} \delta V_{kk} \delta_{ij} + f_{\perp} \delta V_{\langle ij\rangle} \,.\\
    \end{split}
\end{equation}
Finally, after using Eq. \eqref{eq:therm1} and Eq. \eqref{eq:pressureDef} we write the variation of the pressure with respect to the thermodynamic variable as
\begin{equation}\label{eq:pressure}
      \delta P = - T_0 ( P_{\epsilon} \delta \epsilon + P_n \delta n ) \,,
\end{equation}
where we have defined
\begin{align}
\nonumber    P_\epsilon &=n_0 s_{n\epsilon} + (P_0 + \epsilon_0) s_{\epsilon \epsilon}\,,\\
\label{eq:pressureNotation}    P_n &=n_0 s_{n n} + (P_0 + \epsilon_0) s_{n \epsilon}\, .
\end{align}
\section{Dipole-conserving hydrodynamics}\label{sec:hydrodynamics}
In this section, we develop the hydrodynamic framework for dipole-conserving  fluids by applying the entropy current formalism. We construct constitutive relations following a derivative expansion and impose the second law of thermodynamics. First, we consider the leading order hydrodynamics, and then we study first-order corrections in a linearized regime. After having the most general constitutive relations, the transport coefficients will be constrained by the entropy production condition.

\subsection{Gradient expansion}\label{sec:grad}
Following the canonical paradigm of hydrodynamics, we consider the long-wavelength, near-equilibrium dynamics that is governed by the hydrodynamic variables, that is, the densities $n,\epsilon,p_i$ of the conserved charges. Macroscopic currents are then given by local expressions of the conserved densities organized in a systematic derivative expansion.
The explicit form of the currents dubbed as \textit{constitutive relations} is fixed by the symmetries in Eqs. \eqref{eq:conserved}. In writing these constitutive relations a set of unknown parameters will emerge, known as transport coefficients, which are then constrained imposing the laws of thermodynamics, and Onsager relations.

Nonetheless, the non-standard structure the dipole symmetry introduces suggest we should consider the momentum of the system $p_i$ to be of order $\mathcal O(p_i)\sim\mathcal O(\partial_i)^{-1}$, such that $V_{ij}\sim\mathcal O(\partial_i)^{0}$. Therefore, our derivative expansion is defined in terms of the order at which the equations of motion are truncated, e.g. we will refer to $n-$th order hydrodynamics if the set of differential equations is truncated as \footnote{Notice that onshell temporal derivatives will not be independent from spatial gradients, in particular we have the hierarchy $\mathcal{O}(\partial_t) \sim \mathcal{O}(\partial_i)^2$.}
\begin{equation}
    \begin{split}
         \partial_t \epsilon   & = -\partial_i J_\epsilon^i + \mathcal O(\partial_i)^{2n+3} \,, \\
        \partial_t p_i  &= -\partial_j T^{ji} + \mathcal O(\partial_i)^{2n+2}  \,, \\
        \partial_t n  & = - \partial_i \partial_j J^{ij} + \mathcal O(\partial_i)^{2n+3}  \,.
    \end{split}
\end{equation}
\subsection{Zeroth order hydrodynamics}
To start, we derive the leading order constitutive relations for dipole conserving fluids. Our results are a dissipative completion of the theory constructed in Ref. \cite{Grosvenor:2021hkn}, where the zero temperature ideal constitutive relations were found using the Poisson bracket formalism. 
The  local form of the first law can be derived from Eq. \eqref{eq:diffentropy} as follows
\begin{equation}\label{eq:passing} 
\begin{split}
      T \partial_t s  &=  \partial_t \epsilon  - \mu \partial_t n -  F_{ij} \partial_t V_{ij} \,,\\
    & =  \partial_t \epsilon  -  \partial_i \left(F_{ij} \partial_t \left(\frac{p_j}{n}\right)\right)  - \tilde{\mu} \partial_t n   -  V_j \partial_t p_j \,,
\end{split}
\end{equation}
in the last step we have defined the effective chemical potential and velocity
\begin{equation}
\label{eq:notation}
 \tilde{\mu} = \mu - \frac{V_ip_i}{n} \,, \quad V_i = - \frac{\partial_j F_{ji} }{n} \,.
\end{equation}
Using the equations of motion Eqs. \eqref{eq:continuity} it is then possible to recast Eq. \eqref{eq:passing} into a familiar looking equation 
\begin{align}
 \label{eq:familiar}
   T \partial_t s = - \partial_i \mathcal{E}^i 
    + \tilde{\mu} \partial_i \partial_j  J^{ij} + V_i \partial_j T^{ji} \,,
\end{align} 
where we have defined a \textit{shifted energy current}
\begin{equation}\label{eq:shifted}
J_\epsilon^i  = \mathcal{E}^i - F_{ij} \partial_t \left(\frac{p_j}{n}\right)\,.
\end{equation}
Throughout the rest of this work, we will often find it convenient to work with the shifted energy current. 

Using Eq. \eqref{eq:familiar} we can express the entropy production constraint $\partial_t s +\partial_i S^i \geq0$ as 
\begin{equation}
\label{eq:entprodc0}
\partial_i S^i - \frac{1}{T}\partial_i \mathcal{E}^i 
    + \frac{\tilde{\mu}}{T} \partial_i \partial_j  J^{ij} + \frac{V_i}{T} \partial_j T^{ji} \geq 0 \,.
\end{equation}
Thus, after combining the thermodynamic relation Eq. \eqref{eq:pressureDef} with Eq.  \eqref{eq:entprodc0} and a series of tedious algebraic computations that we show in Appendix \ref{sec:nextToLeading}, it is possible to rewrite the constraint as  
\begin{widetext}
\begin{equation}
\begin{gathered}
     \partial_i \Big( S^i - \frac{1}{T} \mathcal{E}^i + \frac{\tilde{\mu}}{T}  \partial_j J^{ij} - \frac{V_j}{T}\Big[  P \delta_{ij}
    +   F_{ik} \partial_j \frac{p_k}{n} +  \partial_k \big(F_{ij} \frac{p_k}{n} - F_{kj} \frac{p_i}{n}\big) - T^{ij} \Big]  \Big) + \Big( \mathcal{E}^i  - (P+\epsilon) V_i \Big) \partial_i (\frac{1}{T}) \\
    + (nV_i - \partial_j J^{ij}) \partial_i (\frac{\tilde{\mu}}{T})  +\Big[P \delta_{ij} + V_i p_j + F_{ik} \partial_j \frac{p_k}{n} +\partial_k \big(F_{ij} \frac{p_k}{n} - F_{kj} \frac{p_i}{n}\big) - T^{ij} \Big]\partial_i (\frac{V_j}{T}) \geq 0 \,. 
\end{gathered}
 \label{eq:master}
\end{equation}
\end{widetext}
Therefore, we conclude that the local version of the second law of thermodynamics will be satisfied provided that the first term in Eq. \eqref{eq:master} vanishes and the remainder is semi-positive definite for arbitrary field configurations. This constraint fixes the zeroth order currents to
\begin{equation}\label{eq:constitutive}
\begin{split}
 J_\epsilon^i &=  (\epsilon + P )V_i - F_{ij} \partial_t(\frac{p_j}{n}) + \alpha \partial_i \frac{1}{T}  \,,\\
  J^{ij} &= - F_{ij}  \,,\\
 T^{ij} &=  P \delta_{ij} + V_i p_j + V_j p_i  + \partial_k F_{ij} \frac{p_k}{n} + F_{ij} V_{kk} \,, \\ 
 S^i &= \frac{1}{T} \mathcal{E}^i  + P \frac{V_i}{T}
    - \frac{\tilde{\mu}}{T}  \partial_j J^{ij} - \frac{V_j}{T} T^{ij} \\ 
    &+ \frac{V_j}{T}  F_{ik} \partial_j \frac{p_k}{n} + \frac{V_j}{T} \partial_k \Big( F_{ij} \frac{p_k}{n} - F_{kj} \frac{p_i}{n} \Big)
\end{split}
\end{equation}
with $\alpha$ a transport coefficient that can be interpreted as the thermal conductivity of the system, satisfying the inequality 
\begin{equation}\label{eq:alpha}
    \alpha \geq 0 \,.
\end{equation}
Actually notice that the stress tensor has the required transformation property under dipole shift Eq. \eqref{eq:transformation2}. In addition, it can be shown that the entropy current reduces to the simple form
\begin{equation}\label{eq:entropy1}
S^i = s V_i +  \frac{\alpha}{T} \partial_i \frac{1}{T} \,.
\end{equation}
It is important to emphasize that both of the terms in the entropy current enter with a single spatial derivative of a hydrodynamic variable. Thus, in dipole conserving systems the limit of ideal hydrodynamics, as constructed in Ref. \cite{Grosvenor_hydrodynamics_2021}, can only be reached by fine-tuning $\alpha=0$ rather than neglecting higher order derivative corrections.  This happens because the lowest order contributions, in our counting scheme, allow for a dissipative transport coefficient. This is at odds with fluids without kinematic constraints.

We now turn our attention to the study of the hydrodynamic modes. To this aim, we consider linearized perturbations around the equilibrium state $(\epsilon_0,n_0,\mathbf p_0=0)$. Therefore, the fluctuations read
\begin{equation}\label{eq:equilibriumH}
    n=n_0 +\delta n, \quad \epsilon=\epsilon_0 + \delta \epsilon, \quad \mathbf p =  \mathbf{\delta p} \,,
\end{equation}
and the corresponding currents \eqref{eq:constitutive} take the form
\begin{equation}\label{eq:linearizedCurrentsZeroth}
\begin{gathered}
         J_\epsilon^i = (\epsilon_0 + P_0) V_i + \alpha \partial_i \frac{1}{T}\,, \\
          J^{ij}  = -F_{ij}\,, \hspace{10px}T^{ij} = (P_0 + \delta P) \delta_{ij} \,.  
\end{gathered}
\end{equation}
In order to solve for the evolution of the hydrodynamic variables, we express all quantities that appear in the above currents in terms of the variations of the conserved densities $\delta n, \delta \epsilon$ and  $\delta \mathbf p$. This is done using the thermodynamic relations Eqs. \eqref{eq:therm1} and  \eqref{eq:pressure}, as well as the definition of the effective velocity Eq. \eqref{eq:notation}. After doing so, the zeroth order hydrodynamic equations of motion become
\begin{equation}\label{eq:linearziedEoms}
    \begin{split}
      \partial_t \delta n -  \bar{f} \nabla^2 \nabla\cdot \mathbf{\delta p} &=0  \,, \\[2.5pt]
       \partial_t \mathbf{\delta p}    - T_0 P_{\epsilon}  \nabla \delta \epsilon -T_0 P_n  \nabla \delta n  &=0 \,, \\
        \partial_t \delta \epsilon +   \alpha s_{ee} \nabla^2 \delta \epsilon + \alpha s_{ne}  \nabla^2 \delta n -\bar{f} \frac{\epsilon_0+ p_0 }{n_0} \nabla^2 \nabla\cdot \mathbf{\delta p}  &= 0
    \end{split}
\end{equation}
where $\bar{f} = n_{0}^{-1} \Big(f_{||}+f_{\perp} \frac{d-1}{d} \Big)$. 

In order to solve the system and find the dispersion relation of the modes, we must Fourier transform the equations with frequencies and momenta ($\omega,k_i$). For this set of equations, the transverse sector ($\mathbf k\cdot\mathbf{\delta p}=0$) contains the non-dispersive mode $\omega_{shear} = 0$.

On the contrary, the longitudinal sector $(\mathbf k\times \mathbf{\delta p}= 0,\delta n, \delta \epsilon)$ is determined by the characteristic polynomial
\begin{equation}\label{eq:polynomial0}
    \left(\frac{\omega}{\omega_0}\right)^3 + i a_2 \left(\frac{\omega}{\omega_0}\right)^2   - \frac{\omega}{\omega_0} - i a_1= 0
\end{equation}
with $\omega_0 = \sqrt{a_0}k^2$, and 
\begin{equation}
\begin{split}
 a_0 &= - T_0 \bar{f} n^{-1}_0 \big[ n_0 P_{n}+(\epsilon_0 + P_0)P_{\epsilon} \big] \,, \\
    a_1 &= -a_0^{-\frac{3}{2}}\alpha T_0  \bar{f} \big[ s_{n\epsilon}  P_{\epsilon} -   s_{\epsilon \epsilon} P_{n} \big]  \,,\\
     a_2 &=  -a_0^{-\frac{1}{2}}\alpha s_{\epsilon \epsilon}  \,,
    \end{split}
\end{equation}
where the conditions Eqs. \eqref{eq:stability} and \eqref{eq:alpha} imply $a_0 \geq 0$ and $0 \leq a_1 < a_2$. 

The solutions to Eq. \eqref{eq:polynomial0} are
\begin{equation}\label{eq:dispersion0}
    \begin{split}
        \frac{\omega_1}{\omega_0} &= -i\frac{x}{3} \left(\frac{x}{ y}-\frac{y}{x}\right) -\frac{1}{3} i a_2\,,\\
      \frac{\omega_2}{\omega_0} & = \frac{x}{2\sqrt{3}} \left(\frac{x}{ y} + \frac{y}{x}\right) + i\frac{x}{6} \left(\frac{x}{ y}-\frac{y}{x}\right)- i\frac{1}{3}   a_2\,,\\
       \frac{\omega_3}{\omega_0} & = -\frac{x}{2\sqrt{3}} \left(\frac{x}{ y}+\frac{y}{x}\right) + i\frac{x}{6} \left(\frac{x}{ y}-\frac{y}{x}\right)- i\frac{1}{3}   a_2 \,.
    \end{split}
\end{equation}
with
\begin{equation}\label{eq:31}
    \begin{gathered}
        x^2=  3-a_2^2\,, \\
       2y^3= -27  a_1  + 9   a_2 - 2  a_2^3 
        \\
        +3\sqrt{81 a_1^2+6 a_1 a_2 \left(2 a_2^2-9\right)-3 a_2^2+12} \,.
    \end{gathered}
\end{equation}
Notice that all modes have a $k^2$ dependence. 
Actually, it is possible to expand the solutions for small and large value of the thermal conductivity. In the former case, which corresponds to $a_2\ll 1$ and $a_1/a_2$ fixed, the modes read
\begin{equation}
\begin{split}
    \omega_1 &\approx -i \sqrt{a_0} a_1 k^2\,,\\
    \omega_2 & \approx\sqrt{a_0} \left(1 -\frac{1}{2}i(a_2-a_1)\right)k^2\\
    \omega_3 & \approx -\sqrt{a_0} \left(1 +\frac{1}{2}i(a_2-a_1)\right)k^2 \,.
\end{split}
\end{equation}
Whereas in the latter ($a_2\gg 1$ and $a_1/a_2$ fixed) the asymptotic behaviors for of the dispersion relations are
\begin{equation}
\begin{split}
    \omega_1 &\approx \sqrt{a_0}\left( \sqrt{\frac{a_1}{a_2}} - i\frac{a_2-a_1}{2a_2^2} \right)k^2\,,\\
    \omega_2 & \approx -i\sqrt{a_0}\left( a_2 - \frac{a_2-a_1}{a_2^2}\right)k^2\,,\\
    \omega_3 & \approx -\sqrt{a_0}\left(  \sqrt{\frac{a_1}{a_2}} + i\frac{a_2-a_1}{2a_2^2}\right)k^2 \,.
    \end{split}
\end{equation}

\begin{center}
    \begin{figure*}[t!]
        \centering
\includegraphics[width=0.66\columnwidth]{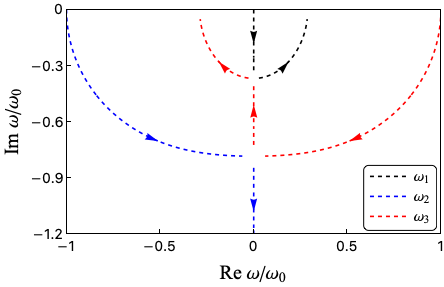}
\includegraphics[width=0.66\columnwidth]{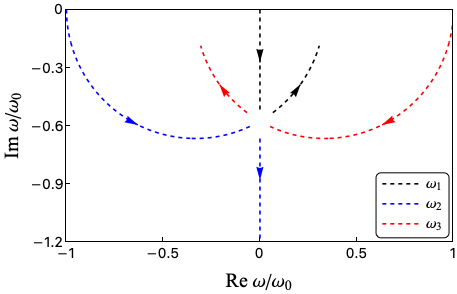}
\includegraphics[width=0.66\columnwidth]{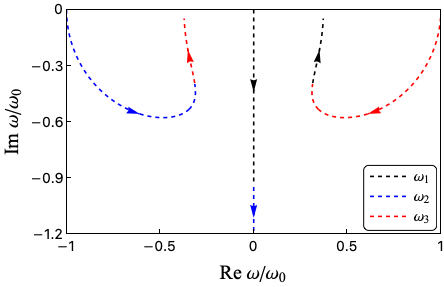}
        \caption{Trajectories of the longitudinal modes in the frequency complex plane as a function of the thermal conductivity at fixed momentum with arrows indicating the direction of increase in thermal conductivity. \textbf{Left:} Frequencies as a function of the thermal conductivity for $a_1/a_2=0.75(a_1/a_2)_c$ . \textbf{Middle:} Longitudinal modes in the critical regime $(a_1/a_2)_c$, the collision of the modes happens at $a_2=\sqrt{3}$. \textbf{Right:} Trajectories in Regime II for  $a_1/a_2=1.25(a_1/a_2)_c$.  }
        \label{fig:fig1}
    \end{figure*}
\end{center}

The full dependence of the modes as a function of the adimensional thermal conductivity $a_2$  at fixed $a_1/a_2$ is shown in Fig. \ref{fig:fig1}. We notice the existence of two qualitatively distinct regimes corresponding to $a_1/a_2<1/9$ (Regime I) and $a_1/a_2>1/9$ (Regime II). Actually, at the critical regime $(a_1/a_2)_c = 1/9$ there is a point for which the three modes are equal and purely imaginary. This situation is shown in the middle plot in Fig. \ref{fig:fig1}. The main difference between Regimes I and II is the existence of a window in the parameter space where the three modes are purely imaginary (Regime I), whereas in Regime II there will always be two modes with a non-vanishing real part. 
Finally, it is worth to point out that the thermodynamic constraints $a_1/a_2<1$ and $\alpha\geq0$ are enough to guarantee that the imaginary part of the modes is negative, and no linear instability will occur.

\subsection{First order hydrodynamics}\label{eq:dissipativeHydrodynamics}

In the previous section, we have shown the existence of one dissipative transport coefficient $\alpha$ that controls how longitudinal fluctuations diffuse in the system. However, the shear mode $\omega_{shear}=0$ remained insensitive to the thermal conductivity. Therefore, at that level of the derivative expansion, transverse fluctuations will not diffuse. Although we may think this fact is reminiscent of the fractonic nature of the system, in this section we will prove that this is not the case. Actually, the first order transport coefficients will introduce transverse contributions to the next to leading order hydrodynamic equations of motion and predict a subdiffusive shear mode.

To proceed with the analysis, we decompose the currents into the zeroth and first order contributions according to the derivative counting scheme 
\begin{equation}
\begin{split}\label{eq:decomposition}
    & J^{ij}=J^{ij}_{0} + J^{ij}_{1}   , \quad \mathcal{E}^{i}=\mathcal{E}^i_{0} + \mathcal{E}^i_{1}\,, \\
   & T^{ij} = T^{ij}_{0} + T^{ij}_{1}  , \quad  S^{i}=S^i_{0} + S^i_{1}  \,,
    \end{split}
\end{equation}
and plug the decomposition Eq. \eqref{eq:decomposition} into Eq. \eqref{eq:passing} and cancel out the lower order terms (as these satisfy the second law provided that $\alpha \geq 0$). Then, the second law requires that 
\begin{equation}\label{eq:2ndlawD}
    \partial_i S^i_{1} - \frac{1}{T}\partial_i \mathcal{E}^i_{1}
    + \frac{\mu}{T} \partial_i \partial_j  J^{ij}_{1} + \frac{V_i}{T} \partial_j T^{ji}_{1} \geq 0.
\end{equation}
Note that for fluctuations around Eq. \eqref{eq:equilibriumH} we have that $\tilde{\mu} = \mu$ and $J^i_{\epsilon} = \mathcal E^i$ (see Eqs. \eqref{eq:notation} and \eqref{eq:shifted}). Since our goal is to identify the most general constitutive relations consistent with the above inequality, we find it convenient to rewrite Eq. \eqref{eq:2ndlawD} as follows
   \begin{equation}\label{eq:diss}
   \begin{gathered}
  \partial_i \Big(S^i_{1}  - \frac{1}{T}  \mathcal{E}^i_{1} + \frac{\mu}{T}  \partial_j J_{1}^{ij} -  \partial_j ( \frac{\mu}{T} ) J_{1}^{ij} + \frac{V_j}{T} T_{1}^{ij}  \Big) \\
 + J_{1}^{ij} \partial_i \partial_j  \frac{\mu}{T}  +  \mathcal{E}_{1}^i \partial_i \frac{1}{T} - T_{1}^{ij} \partial_i \frac{V_j}{T} \geq 0\,.
   \end{gathered}
 \end{equation}
Ignoring non-linearities, it is possible to express the energy current without loss of generality as $\mathcal{E}_1^i = \partial_j E^{ji}$. With that ansatz at hand, the entropy production constraint can be written as 
   \begin{equation}\label{eq:diss2}
   \begin{gathered}
  \partial_i \Big(S^i_{1}  - \frac{1}{T}  \partial_j E^{ji} + \partial_j \frac{1}{T} E^{ij} + \frac{\mu}{T}  \partial_j J_{1}^{ij} -  \partial_j  \frac{\mu}{T}  J_{1}^{ij} + \frac{V_j}{T} T_{1}^{ij}  \Big) \\
 + J_{1}^{ij} \partial_i \partial_j  \frac{\mu}{T}  -  E^{ij} \partial_i \partial_j \frac{1}{T} - T_{1}^{ij} \partial_i \frac{V_j}{T} \geq 0 \,.
   \end{gathered}
 \end{equation}
Then, we set the first order correction to the entropy current to be
\begin{equation}
S^i_{1}  = \frac{1}{T}  \partial_j E^{ji} - \partial_j \frac{1}{T} E^{ij} - \frac{\mu}{T}  \partial_j J_{1}^{ij} +  \partial_j  \frac{\mu}{T}  J_{1}^{ij} - \frac{V_j}{T} T_{1}^{ij} \,,
\end{equation}
such that the first term in Eq. \eqref{eq:diss2} vanishes. Therefore, the problem is reduced to finding the proper constitutive relations such that the leftover is semi-positive definite. In fact, the most general form for the currents that will allow a positive production read
\begin{equation}\label{eq:macroscopic}
\begin{split}
    J^{ij}_1 &=  \Big( \frac{\gamma_{||} }{T_0} \partial_k V_k + \sigma_{||} \nabla^2 \frac{\mu}{T} + \beta_{||} \nabla^2 \frac{1}{T} \Big) \delta_{ij} \\
    &+  \frac{\gamma_{\perp} }{T_0} \partial_{\langle i} V_{j \rangle}  + \sigma_{\perp} \partial_{\langle i} \partial_{j \rangle} \frac{\mu}{T}  + \beta_{\perp} \partial_{\langle i} \partial_{j \rangle} \frac{1}{T} \,, \\
     - T^{ij}_1 &= \Big( \frac{\zeta}{T_0} \partial_k V_k + \bar{\gamma}_{||} \nabla^2 \frac{\mu}{T} + \alpha_{||} \nabla^2 \frac{1}{T} \Big) \delta_{ij} \\
    &+ \frac{\eta}{T_0}\partial_{\langle i} V_{j \rangle}  + \bar{\gamma}_{\perp} \partial_{\langle i} \partial_{j \rangle} \frac{\mu}{T}  + \alpha_{\perp} \partial_{\langle i} \partial_{j \rangle} \frac{1}{T} \,,\\
     - E^{ij} & = \Big( \frac{\bar{\alpha}_{||}}{T_0} \partial_k V_k + \bar{\beta}_{||} \nabla^2 \frac{\mu}{T} + \kappa_{||} \nabla^2 \frac{1}{T} \Big) \delta_{ij} \\
    &+ \frac{\bar{\alpha}_{\perp} }{T_0} \partial_{\langle i} V_{j \rangle}  + \bar{\beta}_{\perp} \partial_{\langle i} \partial_{j \rangle} \frac{\mu}{T}  + \kappa_{\perp} \partial_{\langle i} \partial_{j \rangle} \frac{1}{T} \,.
    \end{split}
\end{equation}
Where we have introduced a set of 18 dissipative transport coefficients. In particular, Onsager reciprocity reduces the number of off-diagonal coefficients if time-reversal invariance is imposed 
\begin{equation}\label{eq:onsager}
    \bar{\alpha}_{||(\perp)} =  \alpha_{||(\perp)}\,, \quad \bar{\beta}_{||(\perp)} =  \beta_{||(\perp)}\,,  \quad \bar{\gamma}_{||(\perp)} =  \gamma_{||(\perp)}\,.
\end{equation}

Moreover, the entropy production constraint Eq. \eqref{eq:diss2} can be written in a compact matrix form
\begin{equation}\label{eq:matrixInequality}
    \bm{x}^{\intercal} \mathcal{A}_{\bm{||}} \bm{x} + \bm{y}^{\intercal}\mathcal{A}_{\bm{\perp}} \bm{y} \geq 0 \,,
\end{equation}
with
\begin{equation}
\begin{gathered}
    \mathcal{A}_{\bm{||}} = \begin{pmatrix}
\zeta & \gamma_{||} & \alpha_{||} \\
\gamma_{||} & \sigma_{||} & \beta_{||}\\
\alpha_{||} & \beta_{||} & \kappa_{||}
\end{pmatrix}, \quad    A_{\bm{\perp}} = \begin{pmatrix}
\eta & \gamma_{\perp} & \alpha_{\perp} \\
\gamma_{\perp} & \sigma_{\perp} & \beta_{\perp}\\
\alpha_{\perp} & \beta_{\perp} & \kappa_{\perp}
\end{pmatrix}  \,, \\
\bm{x} = \begin{pmatrix}
   \partial_i \frac{V_i}{T} \\ \nabla^2 \frac{\mu}{T} \\ \nabla^2 \frac{1}{T} \end{pmatrix}\,, \hspace{5px}\bm{y} = \begin{pmatrix}
   \partial_{\langle i} \frac{V_{i \rangle}}{T} \\ \partial_{\langle i} \partial_{j \rangle} \frac{\mu}{T} \\ \partial_{\langle i} \partial_{j \rangle} \frac{1}{T}
\end{pmatrix} \,.
\end{gathered}
\end{equation}
Since the two contributions are independent, the second law is then imposed by requiring that matrices $\mathcal{A}_{\bm{||}}$ and $\mathcal{A}_{\bm{\perp}}$ are both semi-positive definite. This poses constraints on the transport coefficients, which are summarized below. 

In total, we have found 12 independent transport coefficients that we classify in two distinct categories. The first category involves the diagonal coefficients $(\zeta,\eta,\sigma_\perp,\sigma_{||},\kappa_\perp,\kappa_{||})$ satisfying a positivity constraint
\begin{equation}\label{eq:viscosities}
    \zeta \,, \eta\,,  \sigma_{\perp} \,, \sigma_{||}\,, \kappa_{\perp}\,, \kappa_{||} \ \geq 0\,.
\end{equation}
On the other hand, the second category consists of the off-diagonal terms $(\alpha_\perp,\alpha_{||},\beta_\perp,\beta_{||},\gamma_\perp,\gamma_{||})$ obeying inequalities with the coefficients of the previous group
\begin{equation}\label{eq:coefficients}
\begin{gathered}
     \alpha^2_{\perp} \leq \sigma_{\perp} \kappa_{\perp}\,, \hspace{10px}\alpha^2_{||} \leq \sigma_{||} \kappa_{||}\,, \hspace{10px}   \beta^2_{\perp} \leq \eta \kappa_{\perp}, \\
     \beta^2_{||} \leq \zeta \kappa_{||}\,, \hspace{10px}
    \gamma^2_{\perp} \leq \eta \sigma_{\perp}\,, \hspace{10px}  \gamma^2_{||} \leq \zeta \sigma_{||}\,.
    \end{gathered}
\end{equation}
The distinction has been motivated by the fact that the value of the off-diagonal transport coefficients is bounded from above by the diagonal ones. The last constraint from semi-positivity corresponds with the positive determinant condition 
\begin{equation}
\begin{split}
       \zeta(\sigma_{||}\kappa_{||}-\beta^2_{||})-\kappa_{||}\gamma^2_{||}
        -\sigma_{||} \alpha^2_{||}\geq 0\,, \\
               \eta(\sigma_{\perp}\kappa_{\perp}-\beta^2_{\perp})-\kappa_{\perp} \gamma^2_{\perp}
               -\sigma_{\perp} \alpha^2_{\perp} \geq 0\,.
               \end{split}
\end{equation}
Having the corrections to the zeroth order hydrodynamic constitutive currents, we can plug them into the conservation equations to obtain the first order hydrodynamic equations of motion. In particular, they read
\begin{equation}\label{eq:eoms1}
    \begin{gathered}
\partial_t \delta n +  j_n \nabla^4 \delta n- (\bar{f} - j_v \nabla^2)\nabla^2(\nabla\cdot \mathbf{\delta p})  + j_e \nabla^4 \delta \epsilon  = 0\,, \\[2.5pt]
\partial_t  \mathbf{\delta p} + t_{v_{||}} \nabla^2 \nabla ( \nabla\cdot \mathbf{\delta p}) +  t_{v_{\perp}} \nabla^4 \mathbf{\delta p} \\
- (T_0 P_n  - t_n  \nabla^2) \nabla \delta n  - (T_0 P_{\epsilon}  -  t_e \nabla^2 ) \nabla \delta \epsilon 
  = 0\,, \\[2.5pt]
\partial_t \epsilon +  (  \alpha s_{ee} +  e_e \nabla^2)\nabla^2 \delta \epsilon + (\alpha s_{ne} + e_n \nabla^2)\nabla^2 \delta n\\
-\left(\bar{f} \frac{\epsilon_0+ p_0 }{n_0}  -e_v \nabla^2\right)\nabla^2( \nabla\cdot \mathbf{\delta p}   )
   = 0\,.
    \end{gathered}
    \end{equation}    
For a detailed derivation of the equations and the relation of the parameters shown in Eqs. \eqref{eq:eoms1} with the transport coefficients in Eqs. \eqref{eq:macroscopic}, we refer the reader to the Appendix \ref{sec:dissipativeCorrections}. The main output of the first order approach is the conversion of the non-dispersive shear mode into a subdiffusive one
\begin{equation}\label{eq:shear}
   \omega_{shear} = -i \eta \frac{f_\perp}{T_0n^2_0} k^4\,.
\end{equation}
On the other hand, the longitudinal modes are not strongly affected by the first order corrections, since their contribution enters at higher order in momentum. In fact, the characteristic polynomial in this case takes the same form as in Eq. \eqref{eq:polynomial0}
\begin{equation}\label{eq:polynomial1}
    \left(\frac{\omega}{\bar{\omega}_0}\right)^3 + i \bar{a}_2 \left(\frac{\omega}{\bar{\omega}_0}\right)^2   - \frac{\omega}{\bar{\omega}_0} - i \bar{a}_1= 0
\end{equation}
where $\bar{\omega}_0 = \sqrt{\bar{a}_0} k^2$ and
\begin{equation}
\begin{split}
\bar{a}_0 &= (a_0 + b_0 k^2) +\mathcal O (k^4)\,, \\
    \bar{a}_1 &= \bar{a}^{-\frac{3}{2}}_0 (a^{\frac{3}{2}}_0 a_1 + b_1 k^2) +\mathcal O (k^4)\,, \\
    \bar{a}_2 &=  \bar{a}^{-\frac{1}{2}}_0 (a^{\frac{1}{2}}_0 a_2 + b_2 k^2)+\mathcal O (k^4)\,.
    \end{split}
\end{equation}
with $b_0, b_1$ and $b_2$ derived in Appendix \ref{sec:dissipativeCorrections}, and shown in Eq. \eqref{eq:bsDef}. Actually, the solutions Eqs. \eqref{eq:dispersion0} still apply, once we substitute $a_1 \rightarrow \bar{a}_1$ and $a_2 \rightarrow \bar{a}_2$ in the Eqs. \eqref{eq:31}.

\section{Discussion}\label{sec:discussion}
We have presented a hydrodynamic theory of isotropic dipole-conserving fluids up to the first order in the derivative expansion. Our construction gives universal lessons about systems with constrained dynamics. 

First, we have shown how to consistently implement the kinematic constraints of dipole-conserving fluids at the level of equilibrium thermodynamics. This thermodynamic state is not the same as for conventional fluids, but requires an additional tensor quantity controlling the flux of dipoles. Secondly, we elucidated the derivative expansion around this equilibrium state and constructed hydrodynamic constitutive relations. Peculiarly, we have found that dipole-conserving fluids are dissipative even if the equations of motion are truncated at the lowest non-trivial order in the derivative expansion.

Our theory gives a conceptually crisp finite temperature description of systems, which preserve charge and its dipole moment, energy and momentum. As a result, it completes previous studies that have given partial answers to this problem for systems without momentum conservation \cite{Gromov_hydro_2020}, without energy conservation \cite{GloriosoLucas22} or without dissipative effects \cite{Grosvenor_hydrodynamics_2021}. 

Using the linear response theory, we have found and classified new $1+12$ transport coefficients that can, in principle, be used as experimental signatures of fracton fluids. In particular, we find non-trivial corrections to both, the transverse and longitudinal modes studied in Ref. \cite{Grosvenor_hydrodynamics_2021} for ideal fracton fluids. One dissipative coefficient, which we identify with the thermal conductivity, contributes to the diffusive transport of longitudinal perturbations. The shear mode, which is non-dispersive at the lowest order in derivatives, becomes subdiffusive after the inclusion of the first order corrections.

The generalization of our formalism to generic multipole-conserving systems is straightforward. In fact, we should expect the real part of the longitudinal modes and the shear mode to acquire a higher exponent in momentum. However, the imaginary part of the longitudinal modes should still have the $-ik^2$ scaling due to its link to the thermal conductivity. Although we have proven a consistent thermodynamic description of the system, its origin from a microscopic perspective is obscure since $\int dx V_{ij}$ does not correspond to any conserved operator. Therefore, a statistical mechanics picture of the thermodynamic theory proposed here is still an open problem.

Finally, let us comment on the non-equilibrium universality classes recently proposed  in \cite{Gromov_hydro_2020,GloriosoLucas22}. In particular, the authors of Ref. \cite{GloriosoLucas22} have shown that dipole-conserving fluids at rest become unstable against thermal fluctuations in less than four spatial dimensions. However, the dipole-conserving models considered there do not assume energy conservation and consequently display subdiffusive scaling of the hydrodynamic modes. Thus, we suspect that including energy conservation changes the universality class and will render the system stable against thermal fluctuations in three spatial dimensions. It would be interesting to investigate this point in greater detail.

\begin{acknowledgments}
A.G. and P.S. have been supported, in part, by the Deutsche Forschungsgemeinschaft through the cluster of excellence ct.qmat (Exzellenzcluster 2147, Project No. 390858490) and the Polish National Science Centre (NCN), Sonata Bis, under Grant No. 2019/34/E/ST3/00405. F.P.-B. has received funding from the Norwegian Financial Mechanism 2014-2021 via the NCN, POLS, under Grant No. 2020/37/K/ST3/03390. P.S. thanks the University of Murcia for hospitality and support from the Talento program.
\end{acknowledgments}

\appendix
\section{Leading order hydrodynamics}
\subsection{Entropy current}
\label{sec:nextToLeading}
In this Appendix we show the algebraic manipulations that were used when going from Eq.  \eqref{eq:entprodc0} to Eq. \eqref{eq:master} explicitly. 

We start by rearranging Eq. \eqref{eq:entprodc0} as a total derivative minus the extra terms
\begin{equation}\label{eq:step1}
\begin{split}
0 &\leq \partial_i \Big( S^i - \frac{1}{T} \mathcal{E}^i 
    + \frac{\tilde{\mu}}{T}  \partial_j  J^{ij}  +\frac{V_j}{T} T^{ij} \Big) \\
    &+ \mathcal{E}^i \partial_i \frac{1}{T} - \partial_j  J^{ij} \partial_i \frac{\tilde{\mu}}{T} -  T^{ji} \partial_i \frac{V_j}{T} \,.
    \end{split}
\end{equation}
Driven by intuition from the conventional hydrodynamics, we now incorporate pressure into our construction by adding zero $0 = -\partial_i (P \frac{V_i}{T}) + P \partial_i \frac{V_i}{T}+\frac{V_i}{T} \partial_i P $ such that \eqref{eq:step1} becomes
\begin{equation}\label{eq:step2}
\begin{split}
0 &\leq  \partial_i \Big( S^i - \frac{1}{T} \mathcal{E}^i - P \frac{V_i}{T}
    + \frac{\tilde{\mu}}{T}  \partial_j  J^{ij}  +\frac{V_j}{T} T^{ij} \Big) \\
    &+ \mathcal{E}^i \partial_i \frac{1}{T} - \partial_j  J^{ij} \partial_i \frac{\tilde{\mu}}{T} +  (P\delta_{ij} - T^{ji}) \partial_i \frac{V_j}{T}  + \frac{V_i}{T} \partial_i P \,.
    \end{split}
\end{equation}
Using the definition of pressure Eq. \eqref{eq:pressureDef} it is possible to evaluate the gradient 
\begin{equation}\label{eq:pres}
    \partial_i P = n \partial_i \mu +  s \partial_i T - F_{jk} \partial_i V_{jk} \,.
\end{equation}
We substitute $\mu = \tilde{\mu} + n^{-1} V_i p_i$ and rewrite the last term in Eq. \eqref{eq:pres} obtaining
\begin{equation}
    \partial_i P = n \partial_i \tilde{\mu} + s\partial_i T + n \partial_i (\frac{V_j p_j}{n})  - F_{jk} \partial_i \partial_j \frac{p_k}{n} \,.
\end{equation}
Let us now express the last term as a total derivative  
\begin{equation}
\begin{split}
    \partial_i P &= n \partial_i \tilde{\mu} + s\partial_i T + n \partial_i (\frac{V_j p_j}{n}) \\& + \partial_j F_{jk} \partial_i \frac{p_k}{n} - \partial_j (F_{jk} \partial_i \frac{p_k}{n}) \,.
    \end{split}
\end{equation}
Using the definition of velocity $V_i = - n^{-1} \partial_j F_{ji}$ and relabelling the dummy indices $k \leftrightarrow j$ in the second-to-last term we find
\begin{equation}\label{eq:pressuregrad}
    \partial_i P = n \partial_i \tilde{\mu} + s\partial_i T + p_j \partial_i V_j - \partial_j (F_{jk} \partial_i \frac{p_k}{n}) \,.
\end{equation}
Thus, the last term in Eq. \eqref{eq:step2} can be written as follows 
\begin{equation}
\begin{split}
        \frac{V_i}{T} \partial_i P &=  \frac{V_i}{T} \Big (n \partial_i \tilde{\mu} + s\partial_i T + p_j \partial_i V_j - \partial_j (F_{jk} \partial_i \frac{p_k}{n}) \Big) \\
        &= nV_i \partial_i \frac{\tilde{\mu}}{T}  -\Big( \tilde{\mu} n V_i  + TsV_i  + V_j p_j V_i \Big) \partial_i \frac{1}{T} \\
        &+  p_j V_i   \partial_i  \frac{V_j}{T} +  F_{jk} \partial_i \frac{p_k}{n} \partial_j \frac{V_i}{T}- \partial_j \Big( \frac{V_i}{T}  F_{jk} \partial_i \frac{p_k}{n}       \Big) \,.
\end{split}
\end{equation}
Relabelling the dummy indices $i \leftrightarrow  j$ in the last two terms and using the definition of pressure Eq. \eqref{eq:pressureDef} we arrive at
\begin{equation}\label{eq:pressureGradient}
\begin{split}
        \frac{V_i}{T} \partial_i P  & = nV_i \partial_i \frac{\tilde{\mu}}{T} - \Big( P + \epsilon \Big) V_i \partial_i \frac{1}{T}  \\
        &+  (p_j V_i +  F_{ik} \partial_j \frac{p_k}{n})  \partial_i \frac{V_j}{T} - \partial_i \Big( \frac{V_j}{T}  F_{ik} \partial_j \frac{p_k}{n}       \Big) \,.
\end{split}
\end{equation}
Plugging \eqref{eq:pressureGradient} back into \eqref{eq:step2} we obtain an expression that closely resembles Eq. \eqref{eq:master}
\begin{widetext}
\begin{equation}
\begin{gathered}
     \partial_i \Big( S^i - \frac{1}{T} \mathcal{E}^i - P \frac{V_i}{T}
    + \frac{\tilde{\mu}}{T}  \partial_j J^{ij} + \frac{V_j}{T} T^{ij} - \frac{V_j}{T}  F_{ik} \partial_j \frac{p_k}{n} \Big) + \Big( \mathcal{E}^i  - (p+\epsilon) V_i \Big) \partial_i (\frac{1}{T}) \\
    + (nV_i - \partial_j J^{ij}) \partial_i (\frac{\tilde{\mu}}{T})  +\Big(P \delta_{ij} + V_i p_j + F_{ik} \partial_j \frac{p_k}{n}  - T^{ij} \Big)\partial_i (\frac{V_j}{T} ) \geq 0 \,.
\end{gathered}
 \label{eq:master2}
\end{equation}
\end{widetext}
It is then straightforward to reach \eqref{eq:master} by adding another zero
\begin{equation}
\begin{split}
    0 &=- \partial_i \Big( \frac{V_j}{T} \partial_k \big(F_{ij} \frac{p_k}{n} - F_{kj} \frac{p_i}{n}\big) \Big) \\&+  \partial_k \big(F_{ij} \frac{p_k}{n} - F_{kj} \frac{p_i}{n}\big) \Big) \partial_i (\frac{V_j}{T}) \,.
    \end{split}
\end{equation}
This final step is necessary in order to obtain a stress tensor that is manifestly symmetric under the exchange of indices. 
\section{Dissipative corrections}\label{sec:dissipativeCorrections}
\subsection{Fluid data classification}
We provide a classification of the independent fluid variables organized according to their transformations under rotations. To this aim we decompose the symmetric tensor $V_{ij}$ as follows 
\begin{equation}
    V_{ij} = \sigma^{ij} + \frac{\delta_{ij}}{d}\theta 
\end{equation}
where we have defined a transverse tensor $\sigma^{ij}$ and a scalar $\theta$ satisfying 
\begin{equation}
    \sigma^{ij} = \partial_{\langle i} \frac{p_{j\rangle}}{n}\,, \hspace{5px}\theta=\partial_{ i} \frac{p_{i}}{n} \,.
\end{equation}
One may then construct independent structures order by order in the gradient expansion according to the power counting scheme established in \ref{sec:grad}. In table \ref{tab:1} we present a list of the onshell independent linear terms up to the first order in the derivative expansion.
\begin{center}
\begin{figure}
 \label{tab:1} \caption{Classification of the onshell independent linear data up to the first order in the derivative expansion. \vspace{3pt}} 
\begin{tabular}{||c|>{\centering\arraybackslash}p{2cm}|>{\centering\arraybackslash}p{2.5cm}|>{\centering\arraybackslash}p{2.5cm}||} 
\hline
 Order   & Scalars   & Vectors  &  Tensors \\     
 \hline\hline 
    $\mathcal{O}(0)$ & $n\,, \epsilon\,, \theta$ & $ \partial_i n\,, \partial_i \epsilon\,, \noindent \vspace{0.5ex} \linebreak$ $\partial_j \sigma^{ij}, \partial_i \theta \noindent\vspace{1ex}\noindent$ & $\delta_{ij}\,, \sigma^{ij}$ \\     
 \hline
 $\mathcal{O}(1)$ & $\nabla^2 n\,, \nabla^2 \epsilon\,, \noindent \vspace{0.5ex} \linebreak$ $\nabla^2 \theta$ & $\partial_i \nabla^2 n\,, \partial_i \nabla^2 \epsilon\,, \noindent \vspace{0.5ex} \linebreak$ $\nabla^2 \partial_j \sigma^{ij}, \partial_i \nabla^2 \theta$ &    $\partial_{\langle i} \partial_{j \rangle } n\,, \partial_{\langle i} \partial_{j \rangle } \epsilon\,, \vspace{0.5ex}\noindent\linebreak$ $\partial_{\langle i} \partial_{j \rangle } \theta\,, \nabla^2 \sigma^{ij}$   \\ [1ex]
 \hline
\end{tabular}
\end{figure}
 \end{center}

\subsection{Dissipative currents and gradient expansion}
Let us now consider the most general form of the first order (linearized around the equilibrium state Eq. \eqref{eq:equilibriumH}) corrections to the currents, written on the basis of the derivative expansion. 
These are given by (see Table \ref{tab:1})
\begin{equation}
\begin{split}\label{eq:macro}
    J_{1}^{ij} &=\Big(j_{n_1}\nabla^2 n + j_{e_1}\nabla^2\epsilon + j_{v_1} \nabla^2\theta  \Big)\delta^{ij}+ j_{v_2} \nabla^2\sigma^{ij}\\
& + j_{n_2} \partial_{\langle i}\partial_{j\rangle}n + j_{e_2}\partial_{\langle i}\partial_{j\rangle}\epsilon + j_{v_3}\partial_{\langle i}\partial_{j\rangle}\theta \,, \\[2.5pt]
     T_{1}^{ij} &=\Big(t_{n_1}\nabla^2 n + t_{e_1}\nabla^2\epsilon + t_{v_1} \nabla^2\theta  \Big)\delta^{ij}+ t_{v_2} \nabla^2\sigma^{ij}\\
& + t_{n_2} \partial_{\langle i}\partial_{j\rangle}n + t_{e_2}\partial_{\langle i}\partial_{j\rangle}\epsilon + t_{v_3}\partial_{\langle i}\partial_{j\rangle}\theta  \,,  \\[2.5pt]
    \mathcal{E}^i_{1} &=e_{n}\partial_i \nabla^2 n + e_{e}\partial_i \nabla^2\epsilon + e_{\theta} \partial_i \nabla^2\theta  + e_{\sigma} \nabla^2\partial_j \sigma^{ij}\,.\\
\end{split}
\end{equation}
In writing Eq. \eqref{eq:macro} we have introduced a new set of phenomenological parameters. These can be related to the dissipative transport coefficients presented in Eq. \eqref{eq:macroscopic}. 

To this aim, we express $V_i$ in terms of the variables $\theta$ and $\sigma^{ij}$ using Eq. \eqref{eq:therm1}
\begin{equation}
    F_{ij} = f_{||} \theta \delta_{ij} + f_{\perp} \sigma_{ij} \rightarrow V_i = -n^{-1}_0 \Big( f_{||} \partial_i \theta + f_{\perp} \partial_j \sigma_{ij} \Big) \,.
\end{equation}
Thus
\begin{equation}
\begin{split}
    \partial_i V_i &= -n^{-1}_0 \Big( f_{||} + f_{\perp}\frac{d-1}{d}  \Big) \nabla^2 \theta\,, \\
    \partial_{\langle i} V_{j\rangle} &= -n^{-1}_0 \Big( f_{||} \partial_{\langle i} \partial_{j \rangle} \theta + f_{\perp}  \nabla^2\sigma_{ij} - f_{\perp}\frac{d-1}{d^2} \nabla^2 \theta \delta_{ij} \Big) \,, \\
    \nabla^2 V_i &= -n^{-1}_0 \Big( f_{||} \partial_i \nabla^2 \theta+ f_{\perp} \nabla^2 \partial_j \sigma^{ij} \Big) \,, \\
    \partial_i \partial_j V_j &= -n^{-1}_0 \Big( f_{||} + f_{\perp}\frac{d-1}{d}  \Big) \partial_i \nabla^2 \theta \,.
    \end{split}
\end{equation}
Hence we see that 
\begin{equation}\label{eq:transport0}
    \begin{split}
        j_{v_1} &= -T^{-1}_0 n^{-1}_0 \Big[  \Big( f_{||} + f_{\perp}\frac{d-1}{d}  \Big) \gamma_{||} - f_{\perp}\frac{d-1}{d^2} \gamma_{\perp} \Big]\,, \\[1pt]
        t_{v_1} &= T^{-1}_0 n^{-1}_0 \Big[  \Big( f_{||} + f_{\perp}\frac{d-1}{d}  \Big) \zeta - f_{\perp}\frac{d-1}{d^2} \eta \Big]\,, \\[1pt]
           e_{\theta} & = T^{-1}_0 n^{-1}_0 \Big[ \big(f_{||} + f_{\perp}\frac{d-1}{d}\big)\big( \alpha_{||} + \alpha_{\perp} \frac{d-1}{d}\big) + \alpha_{\perp} f_{||} \Big] \,, \\[1pt]
        j_{v_2} &= -T^{-1}_0 n^{-1}_0 f_{\perp} \gamma_{\perp} \,, \hspace{10px} t_{v_2} = T^{-1}_0 n^{-1}_0 f_{\perp} \eta  \,, \\[1pt]
        j_{v_3} &= -T^{-1}_0 n^{-1}_0 f_{||} \gamma_{\perp} \,, \hspace{10px} t_{v_3} = T^{-1}_0 n^{-1}_0 f_{||} \eta \,. \\[1pt]
        e_{\sigma} & = T^{-1}_0 n^{-1}_0 f_{\perp} \alpha_{\perp} \,.
    \end{split}
\end{equation}
Even though $j_{v_2}(t_{v_2})$ and $j_{v_3}(t_{v_3})$ are \textit{a priori} independent parameters, they are in fact related via $\frac{j_{v_2}}{ j_{v_3}} = \frac{t_{v_2}}{ t_{v_3}} = \frac{f_{\perp}}{f_{||}}$ by the requirement of the non-negative entropy production Eq. \eqref{eq:diss2}.

Now, we re-express the terms in Eq. \eqref{eq:macroscopic} involving $\delta \frac{1}{T}$ and $\delta \frac{\mu}{T}$ in terms of the variables $\delta \epsilon$ and $\delta n$. Using the thermodynamic relations Eqs. \eqref{eq:therm1} we can identify 
\begin{equation}\label{eq:transport2}
\begin{split}
    j_{n_1} &=  s_{n\epsilon} \beta_{||} -s_{nn} \sigma_{||}\,, \hspace{10px} j_{n_2} =  s_{n\epsilon} \beta_{\perp} -s_{nn} \sigma_{\perp}\,, \\
        j_{e_1} &=  s_{\epsilon \epsilon} \beta_{||} -s_{n\epsilon} \sigma_{||}\,, \hspace{10px} j_{e_2} =  s_{\epsilon \epsilon} \beta_{\perp} -s_{n\epsilon} \sigma_{\perp}\,, \\
            t_{n_1} &=  s_{nn} \gamma_{||} - s_{n\epsilon} \alpha_{||} \,, \hspace{10px} t_{n_2} =  s_{nn} \gamma_{\perp} - s_{n\epsilon} \alpha_{\perp} \,, \\
        t_{e_1} &= s_{n\epsilon} \gamma_{||} - s_{\epsilon \epsilon} \alpha_{||} \,, \hspace{10px} t_{e_2} =  s_{n\epsilon} \gamma_{\perp} - s_{\epsilon \epsilon} \alpha_{\perp} \,. \\
        e_n & = s_{nn} \big( \beta_{||} + \frac{d-1}{d} \beta_{\perp} \big)  - s_{n \epsilon}  \big( \kappa_{||} + \frac{d-1}{d} \kappa_{\perp} \big) \,, \\
        e_{e} & = s_{ee} \big( \kappa_{||} + \frac{d-1}{d} \kappa_{\perp} \big) - s_{n \epsilon} \big( \beta_{||} + \frac{d-1}{d} \beta_{\perp} \big)\,.
    \end{split}
\end{equation}

\subsection{Linearized equations of motion}\label{sec:linearizedeoms}
In this Appendix, we derive the linearized equations of motion with the first order corrections (Eqs. \eqref{eq:eoms1}). This is most easily done in the basis used in Eqs. \eqref{eq:macro}, which are, of course, completely equivalent to Eqs. \eqref{eq:macroscopic} provided that the transport coefficients are identified according to Eqs. \eqref{eq:transport0} and \eqref{eq:transport2}.

It is then straightforward to compute the relevant gradients of the dissipative currents 
\begin{equation}
\begin{split}\label{eq:currentsGradients}
    \partial_i \partial_j J_{1}^{ij} &= \Big( j_{n_1} +\frac{d-1}{d} j_{n_2} \Big) \nabla^4 n + \Big( j_{e_1} +\frac{d-1}{d} j_{e_2} \Big) \nabla^4 \epsilon  \\
    & + \Big( j_{v_1} + \frac{d-1}{d} \big( j_{v_2}+j_{v_3} \big) \Big) \nabla^4 \theta\,, \\[1pt]
     \partial_j T_{1}^{ij} &=\Big(t_{n_1} +  \frac{d-1}{d} t_{n_2} \Big) \partial_i \nabla^2 n + \Big(t_{e_1} +  \frac{d-1}{d} t_{e_2} \Big) \partial_i \nabla^2 \epsilon   \\ 
     &+\Big(t_{v_1} +  \frac{d-2}{d} t_{v_2} + \frac{d-1}{d} t_{v_3}  \Big) \partial_i \nabla^2 \theta  + n_0^{-1} t_{v_2} \nabla^4 \delta p_i \,, \\[1pt]
    \partial_i \mathcal{E}^i_{1} &=e_{n} \nabla^4 n + e_{e} \nabla^4 \epsilon + \Big( e_{\theta} + \frac{d-1}{d}  e_{\sigma} \Big)  \nabla^4 \theta\,.\\
\end{split}
\end{equation}
Thus, we see that the linearized equations of motion up to first order in the derivative expansion are given by Eqs. \eqref{eq:eoms1} with
\begin{equation}\label{eq:eomsDefs}
    \begin{split}
        j_n &= j_{n_1} +\frac{d-1}{d} j_{n_2}\,, \quad
        j_e = j_{e_1} +\frac{d-1}{d} j_{e_2}\,,\\
        j_v &= j_{v_1} + \frac{d-1}{d} \big( j_{v_2}+j_{v_3} \big)\,, \quad  t_{v_{\perp}} =n^{-1}_0 t_{v_2}\,, \\
         t_{v_{||}} &= t_{v_1} +  \frac{d-2}{d} t_{v_2} + \frac{d-1}{d} t_{v_3} \,, \\
         t_n &= t_{n_1} +\frac{d-1}{d} t_{n_2}\,, \quad
        t_e = t_{e_1} +\frac{d-1}{d} t_{e_2}\,,\\
          e_n &= e_{n_1} +\frac{d-1}{d} e_{n_2}\,, \quad
        e_e = e_{e_1} +\frac{d-1}{d} e_{e_2}\,,\\
        e_v &= e_{\theta} + \frac{d-1}{d}  e_{\sigma}\,.
    \end{split}
\end{equation}
Going to Fourier space, we find that the shear mode picks up a subdiffusive contribution Eq. \eqref{eq:shear} while the dispersion relations of the longitudinal modes are now given by the roots of the modified polynomial Eq. \eqref{eq:polynomial1} where
\begin{equation}\label{eq:bsDef}
    \begin{split}
        b_0 & = \alpha  s_{\epsilon \epsilon}\Big(j_n+t_v\Big)  + \bar{f}\Big( t_n+ n^{-1}_0 (p_0 +\epsilon_0) t_e \Big)\,,  \\
    & +T_0 \Big(e_v P_{\epsilon} +  j_v P_n\Big) - \alpha j_e s_{ne} \\[2.5pt]
    b_1 & = \bar{f}  P_{\epsilon} T_0\Big(e_n -n^{-1}_0 (p_0 +\epsilon_0) j_n \Big)  \\
    &- \bar{f}  P_n T_0 \Big(e_e - n^{-1}_0 (p_0 +\epsilon_0)j_e \Big)\\
    & - \alpha s_{n \epsilon} \Big(\bar{f} t_e + j_v P_{\epsilon} T_0\Big) + \alpha s_{\epsilon \epsilon} \Big(\bar{f} t_n + j_v P_n T_0\Big)\,, \\[2.5pt]
        b_2 &= e_e + j_n + t_v \,.
    \end{split}
\end{equation}

\bibliography{apssamp}

\end{document}